\documentclass[journal=jacsat,manuscript=communication]{achemso}

\SectionsOn
\usepackage[version=3]{mhchem}
\usepackage{xcolor}
\usepackage[utf8]{inputenc}
\usepackage[T1]{fontenc}
\usepackage{subcaption}
\usepackage{multicol, blindtext}
\usepackage{comment}
\usepackage{float}
\usepackage{bm}
\usepackage{verbatim}

\newcommand{\SM}[1]{\textcolor{green}{{\bf Sughra: #1 }}}

\newcommand*{\sometext}{Recently, optical bound states in continuum in various passive photonic crystals have been identified and similar structures incorporated with optical gain have been reported to exhibit lasing. However, no explicit control over the type of lasing BIC has been reported. In this work, we utilize all four fundamental BICs related to the lowest energy $\Gamma$-point of a square photonic crystal lattice. We identify the associated topological charges from experimentally obtained dispersions, finite element method simulations, as well as from spherical decomposition method based on the microscopic polarization currents in the photonic crystal plane. By tailoring the  periodicity and the hole diameter of the photonic crystal slab, we selectively bring each of the four BIC resonances to a wavelength regime, where fluorescent IR702 molecules overlaid with the photonic crystal provide sufficient gain for the onset of lasing. We experimentally analyze all four observed lasing BICs by imaging their far-field polarization vortices and their associated topological charges. The results  correspond excellently with the transmission results as well as the simulation results in the absence of gain. Finally, we experimentally  present a case where the lasing signal reveals the coexistence of two BICs with opposite topological charges, resulting in a unique polarization pattern. We  believe our results enable tailoring the properties, such as polarization winding and topological charge of BICs, by \textit{a priori} design and thus pave the way for a more general utilization of their appealing properties. }

\let\oldmaketitle\maketitle
\let\maketitle\relax

\author{Sughra Mohamed}
\affiliation{Institute of Photonics, University of Eastern Finland, P.O. Box 111, FI-80101 Joensuu, Finland}
\altaffiliation{Authors contributed equally to this work}
\author{Jie Wang}
\affiliation{Department of Physics, Fudan University, Shanghai, China}
\altaffiliation{Authors contributed equally to this work}
\author{Heikki Rekola}
\author{Janne Heikkinen}
\author{Benjamin Asamoah}
\affiliation{Institute of Photonics, University of Eastern Finland, P.O. Box 111, FI-80101 Joensuu, Finland}
\author{Lei Shi}
\email{lshi@fudan.edu.cn}
\affiliation{Department of Physics, Fudan University, Shanghai, China}
\author{Tommi K. Hakala}
\email{tommi.hakala@uef.fi}
\affiliation{Institute of Photonics, University of Eastern Finland, P.O. Box 111, FI-80101 Joensuu, Finland}

\title{Topological charge engineering in lasing bound states in continuum}
\abbreviations{topological charge,BIC,lasing,PhC}

\begin{document}
 
\twocolumn[
\begin{@twocolumnfalse}
\oldmaketitle
\begin{abstract}
\sometext
\end{abstract}
\end{@twocolumnfalse}
]

\flushbottom
\maketitle

\thispagestyle{empty}

\section*{Introduction}

Bound states in continuum (BICs) are states that are spatially bounded, but nevetheless coexist in the  same energy range with unbounded (continuum) states \cite{Hsu2016}. Various systems governed by wave-phenomena exhibit BICs, with examples found for instance in quantum mechanics \cite{Wigner, PhysRevA.11.446}, fluids \cite{Ursell, Cobelli_2009}, and electromagnetic waves \cite{Koshelev2019, Hsu2016}. In photonic crystals (PhCs),  early studies found that the symmetry mismatch between mode inside the crystal and the propagating modes in the radiation continuum results in extremely high radiative Q-factors \cite{PhysRevB.62.4204, PhysRevB.63.125107, PhysRevB.65.235112}, a feature which was later associated with BICs \cite{PhysRevLett.100.183902,Sadrieva2019,Koshelev2019}
Such modes are typically located at the high symmetry points of the lattice \cite{PhysRevLett.113.257401,Sadrieva2019,Kitamura2012,ha2018directional}. Recently, another type of BIC not protected by the symmetry of the lattice, was found by continuously tuning the in-plane wavevector \cite{Hsu2013,Doeleman2018,Yu2019}. In this case the (so-called "accidental") BIC is formed due to destructive interference of the all the radiative modes at a particular position of the dispersion \cite{Ni2016,Yang2014}. 

The pioneering work of Soljacic et al. \cite{PhysRevLett.113.257401} associated the existence 
of BICs to the vortex centers of polarization fields in the far field radiation, characterized by the topological charge. This charge is a conserved quantity given by the winding number of polarization around the vortex center. It is exactly the conservation of topological charge that renders BICs robust against imperfections or disorder in the lattice: a slight perturbation in system parameters is not sufficient to alter the topological properties of the system.

Very recently, both symmetry protected as well as accidental BICs were reported to exhibit lasing action \cite{Kodigala2017, Huang1018, PhysRevB.102.045122, Wu2020, ha2018directional}. For symmetry protected BICs, lasing for both TE and TM-like BICs with +1 and -1 topological charges have been observed at the $\Gamma$-point of a square lattice. In each of the above reports, either one or two out of four possible BICs exhibited lasing. Further, no explicit control of the topological charge or the type of BIC 
was demonstrated.
Here, we demonstrate lasing action from each of the four types of BICs in a square lattice of a photonic crystal slab. By tailoring the hole diameter and the  periodicity of the lattice, lasing action from any of the first order BICs can be realized. For each BIC, namely TE+1, TE-1, TM-1 and TM+1 (here the number indicates the topological charge), a donut shape vector beam with specific polarization properties is observed. The expected polarization pattern agrees excellently with the prediction from our simulations. To highlight how the lattice parameters can be utilized for the control of the lasing BICs, we fine tuned the geometry of the photonic crystal to enable lasing from two non degenerate BICs of opposite topological charge, resulting in a strikingly different polarization pattern as compared to previous cases.

\begin{figure}[H]
\centering
\includegraphics[width=\linewidth]{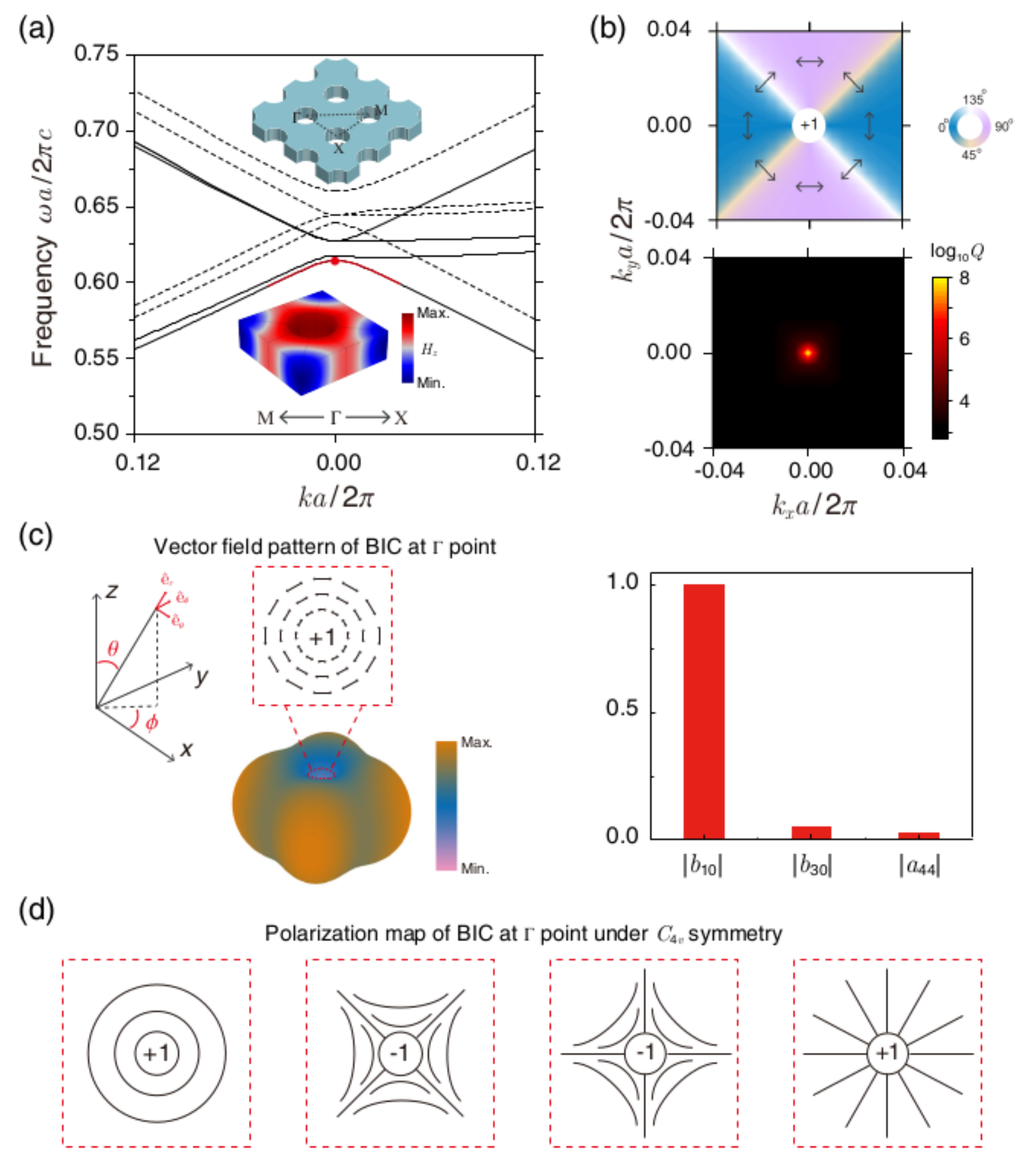}
\caption{(a) Simulated band structure of the 2D PhC slab with square lattice (see also Methods). The bands associated with the TE and TM  are denoted by solid and dashed curves, respectively. Inset: Schematic view of the structure (upper) and electric field distribution of the unit cell (lower). In the simulations, the parameters of the PhC slab are set as follows: refractive index 1.975, thickness 170 nm, periodicity 550 nm, diameter of air hole 250 nm. The environment refractive index is 1.45. (b) Polarization vector distribution as the major axis of polarization states (upper) and calculated radiative quality factor (lower) of the lowest TE band. The band range is denoted by red curve in (a). (c) Reconstructed far-field radiation pattern of the lowest TE-mode at the $\Gamma$-point (marked by red dot in (a)) by multipole decomposition method (left) and the corresponding multipolar composition (right). The vector field distribution very close to the radiation singularity is plotted in red box. (d) Vector field patterns close to BICs with different indices which have covered all sorts of $\Gamma$-point BICs at the lowest bands under C$_{4v}$ symmetry.}
\label{fig11}
\end{figure}

\section*{Theory}
Two methods employed for analyzing BICs are a direct observation of far-field polarization vectors in the momentum space and a recently introduced multipole decomposition method \cite{PhysRevLett.113.257401, chen2019singularities}. While the first method relies on the far field properties of the light and provides a direct observation of topological charges, the second method can be used to calculate the multipolar composition of the near-field polarization currents at the plane of the sample.  
In this work, we will use both methods to study the properties of the observed BICs.

Our structures are two-dimensional (2D) photonic crystal (PhC) slabs with a square array of cylindrical holes (see upper inset in Fig. 1), possessing 4-fold rotational symmetry. The corresponding band structures can be calculated by using a finite-element method (FEM) by Comsol Multiphysics. As shown in Fig. 1, we can find the dispersion relations for TE and TM modes (solid and dashed lines, respectively). We employ the two methods mentioned above to study the symmetry-protected BICs at the $\Gamma$-point and pinpoint all singularities at the lowest bands. Consider first the lowest energy TE mode in Fig. 1(a), whose polarization vector distribution and calculated radiative quality factor Q are plotted in Fig. 1 (b) in the vicinity of the $\Gamma$-point (the band range is also denoted by the red curve in Fig. 1(a)). Notably, the vector field has a singularity with a topological charge $q = +1$ which is given by the winding number of the polarization around the $\Gamma$-point, $ q = \frac{1}{2\pi}\oint_C d \bf{k} \cdot \nabla_k \phi(k) $. Here $\phi(k)$ is the angle of the polarization vector and C is a closed counterclockwise loop around the $\Gamma$-point in k-space \cite{PhysRevLett.113.257401}. The calculated radiative Q factor in Fig. 1 (b) lower panel approaches infinity at $\Gamma$-point, a well known consequence of the symmetry mismatch of such polarization vortex with the radiation fields propagating in free space. 
    The at-$\Gamma$ BICs are antisymmetric solutions to Bloch wave equation. Such modes always fulfill a condition $p(r) = -p(-r)$, where $p$ is the in-plane polarization in the unit cell with a distance $r$  from the inversion center. Thus, for each location $r$ in every unit cell, there exists an equal magnitude but opposite polarization at $-r$. Therefore the radiation in the normal direction of the photonic crystal always experiences destructive interference, resulting in infinite radiative Q-factors.
    
We further analyze the BIC with the multipole decomposition method \cite{PhysRevLett.113.257401, chen2019singularities}. For PhC slab, the multipolar expansion is carried out based on the near-field polarization currents of the unit cell (whose electric field distribution is plotted in lower inset in Fig. 1 (a). Figure 1(c) shows the reconstructed far-field radiation pattern of this $\Gamma$-point BIC and the corresponding multipolar composition (normalized magnitudes of expansion electric $a_{lm}$\ and magnetic $b_{lm}$ coefficients). It can be seen that there exists a radiation singularity along z direction which overlaps with the allowed radiation channel. Meanwhile we can also obtain the vector field distribution very close to this radiation singularity (shown in red box) forming a closed loop with a counterclockwise rotation in agreement with the first method. This $\Gamma$-point state mainly has magnetic dipole part $b_{10}$ whose radiation is intrinsically zero along z direction, thereby leading to the singularity. We confirm again this radiation singularity with the index of +1 according to the singularity index formula, $q=IND={1-|m|}$ . Therefore, the singularity indices of multipolar radiations from multipolar analysis have an equivalent connection with topological charges of BICs. In Fig. 1 (d) we plot all the possible topological charges of BICs for the lowest order TE and TM modes shown in Fig. 1 (a) . Notably, there  are two $\Gamma$-point BICs with two opposite topological charges, +1 and -1 for  the TE and TM modes, respectively. 

\section*{Results}

\begin{figure}[ht]
\centering
\includegraphics[width=1\linewidth]{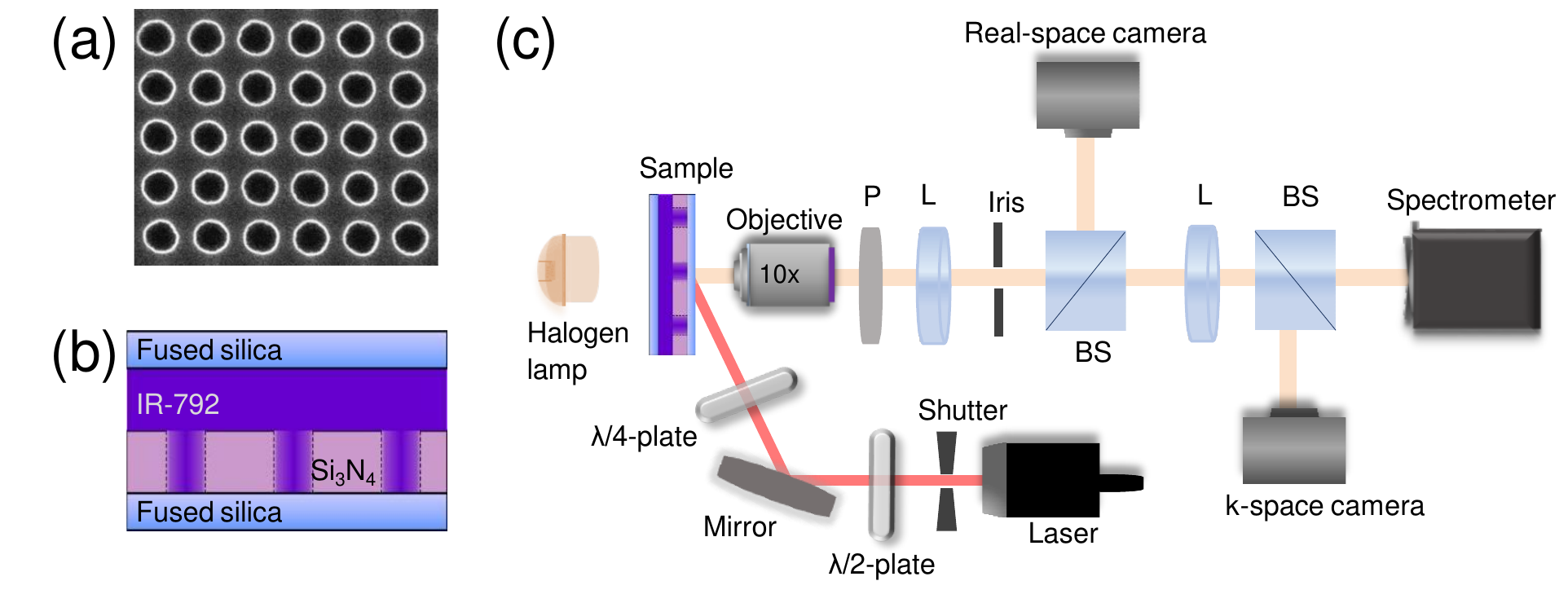}
\caption{(a) SEM image for periodicity of 550 nm. b) Schematic of the sample, consisting of square arrays of holes in silicon-nitride on a fused silica substrate, index matching liquid (and gain medium of (IR-792) for lasing experiments) and a cover glass. c) Schematic of the angle and wavelength  resolved measurement setup for the transmission/lasing experiments. Transmission as well as angle resolved emission spectra are measured by focusing the  back focal plane of the objective lens to the entrance slit of the spectrometer. In the lasing measurements, IR-792 gain medium was inserted between the sample and a cover slip and pumped using circularly polarized 120 fs laser pulses with a central wavelength of 780 nm and 1 kHz repetition rate. The linear rotatable polarizer after the sample was used to analyze the polarization properties of the emission. P = polarizer, L = lens, BS = beam splitter.}
\label{fig1}
\end{figure}

Our sample consists of square lattice of cylindrical holes with depth about 120~nm on 170~nm-thick  Si$_3$N$_4$ layer on a SiO$_2$ substrate, see Fig. 2 (a). Ethylene glycol(ethane-1,2-diol) (EG) and Benzyl alcohol (BA)
mixed in a 1:1 volume ratio was injected on top of the sample to provide index matching with the substrate and sealed with a glass coverslip as seen in Fig. \ref{fig1} (b).

To obtain the band structure, we performed angle and wavelength resolved transmission measurements with a Halogen lamp as a white light source.  The transmitted light from the sample region was spatially selected with an iris. The backfocal plane of the objective lens (10x 0.3NA) was imaged to the spectrometer entrance slit aligned parallel with the y-axis of the sample. 
For lasing measurements, organic dye molecules (IR-792) were added to the EG:BA solution in 25 mM concentration. Optical pumping was performed with a femtosecond laser (120 fs, 1 kHz, 780 nm). The diameter of the pump spot was around 100 $\mu$m  with a 45 degree incidence angle. The  excitation light was circularly polarized, and a  rotatable linear polarizer was used to obtain the polarization properties of the emission.  
The direct k-space imaging was performed by imaging the back focal plane of the objective with a separate camera.

\begin{figure*}[ht]
\centering
  \includegraphics[width=\textwidth]{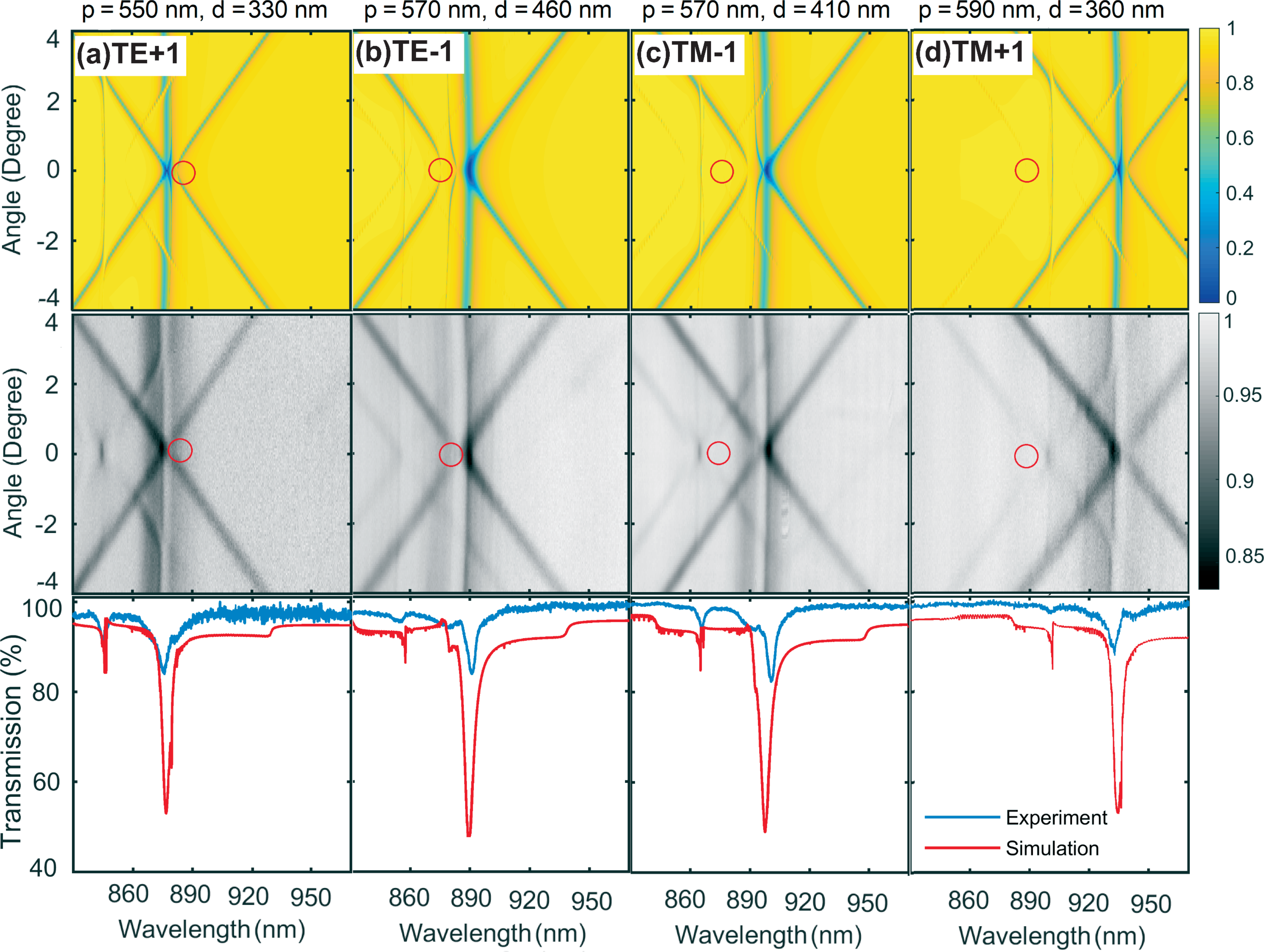}
\caption{FEM simulations and experimentally obtained unpolarized transmissions for 4 samples having different periodicities and hole diameters. The measured spectrum shows the dispersion near the $\Gamma$-point for TE modes (longer wavelength) and TM modes (shorter wavelength). The BIC states obtained from the simulations are indicated by the red circles. In a), the periodicity p = 550 nm and diameter d = 330 nm (TE+1 BIC is at 875 nm). b) p = 570 nm, d = 460 nm TE-1 BIC is at 880 nm. c) p = 570 nm, d = 410 nm TM-1 BIC is at 871 nm. d)  p = 590 nm, d = 360 nm. TM+1 BIC is at 891 nm.}
\label{T-fig}
\end{figure*}
\subsection*{Transmission}
In Fig. \ref{T-fig} the transmission data in $\Gamma -$X direction for 4 samples with different periodicities and hole diameters are presented. The top row shows the FEM simulations, the middle row shows the measured unpolarized transmission spectra, and the bottom row represents transmission spectra under direct incidence. For the sample having p = 550 nm, d = 330 nm in Fig.~3 (a) left column,  the most prominent feature at 875 nm, can be identified as the TE mode. The transmission minimum at around 875 nm 
is due to the doubly degenerate bright mode supported by the lattice, see also Fig. 1 (a). The BIC of interest is located at a slightly higher wavelength (indicated by the red circle), with the transmission approaching 100 \% under direct incidence. The simulations in Fig. \ref{fig11} (a and b) indicate this mode is non-degenerate with a counterclockwise winding of the polarization vector around the $\Gamma$-point indicating topological charge of +1 and a diverging radiative Q-factor at the $\Gamma$-point. The spherical wave decomposition of the near field distributions in Fig \ref{fig11} (c) indicates the mode is mainly composed of the |b$_{10}$| component, resulting in a topological charge of +1. This is in agreement with the farfield polarization pattern seen in Fig \ref{fig11} (b). We also note another, less pronounced feature at around 845 nm, which we identify as TM mode.

   Importantly, the position of the bright mode can be changed from low to the high wavelength side of the diffracted order by tuning the PhC periodicity and the diameter of the holes. To rationalize this, we use simple intuitively clear arguments for 1D photonic crystals \cite{PhysRevB.54.6227}, where the 2 possible solutions to Bloch wave equation are either the symmetric or the antisymmetric mode. At specific filling fraction $f'$ (ratio of thicknesses of dielectric stacks), the modes will be degenerate, and at any other filling fraction the 2 modes will have different energies characterized by the photonic band gap. With values $f < f'$ the higher energy mode becomes the symmetric (bright) mode and with values  $ (f > f')$, the lower energy mode becomes symmetric. Similarly, in 2D lattices, the diameter of the hole governs which side of the diffracted order the bright mode resides. By a proper choice of the diameter, the spectral separation of the BIC of interest and the bright mode can increased. We note that for small diameters, one BIC related to the TE mode is located  at the long wavelength side of the diffracted order while the other BIC with the opposite topological charge and the bright mode are located at the short wavelength side. The bright mode location  switches to the long wavelength side when the diameter is increased, see Supporting information.
   

In Fig. \ref{T-fig} (b) the parameters are p = 570 nm, d = 460 nm. By increasing the periodicity by 20 nm and diameter by 130 nm from the previous case in Fig. 3 (a),  we increase the wavelength of the TE modes by approximately 11 nm 
and switch the bright mode to the high wavelength side of the diffracted order. The BIC nature as well as topological charge of -1 of the low wavelength mode is verified by the simulations which are shown in Fig. 5 (a) for all the remaining BICs. By reducing the diameter to 410 nm, Fig. \ref{T-fig} (c), the features in the dispersion are only slightly redshifted, but this minor variation has a pronounced effect on system response in stimulated emission regime as will be shown below. The red circle indicates the TM-1 BIC, whose resonance wavelength and topological charge are estimated to be 891 nm and -1 from the simulations. 
Finally, in Fig. \ref{T-fig} (d), with parameters p = 590 nm, d = 360 nm, the TM+1 BIC is redshifted sufficiently to overlap with the gain profile of IR-792 and produce lasing action. 

\subsection*{4 different types of lasing BICs}

\begin{figure*}[ht]
\centering
  \includegraphics[width=\textwidth]{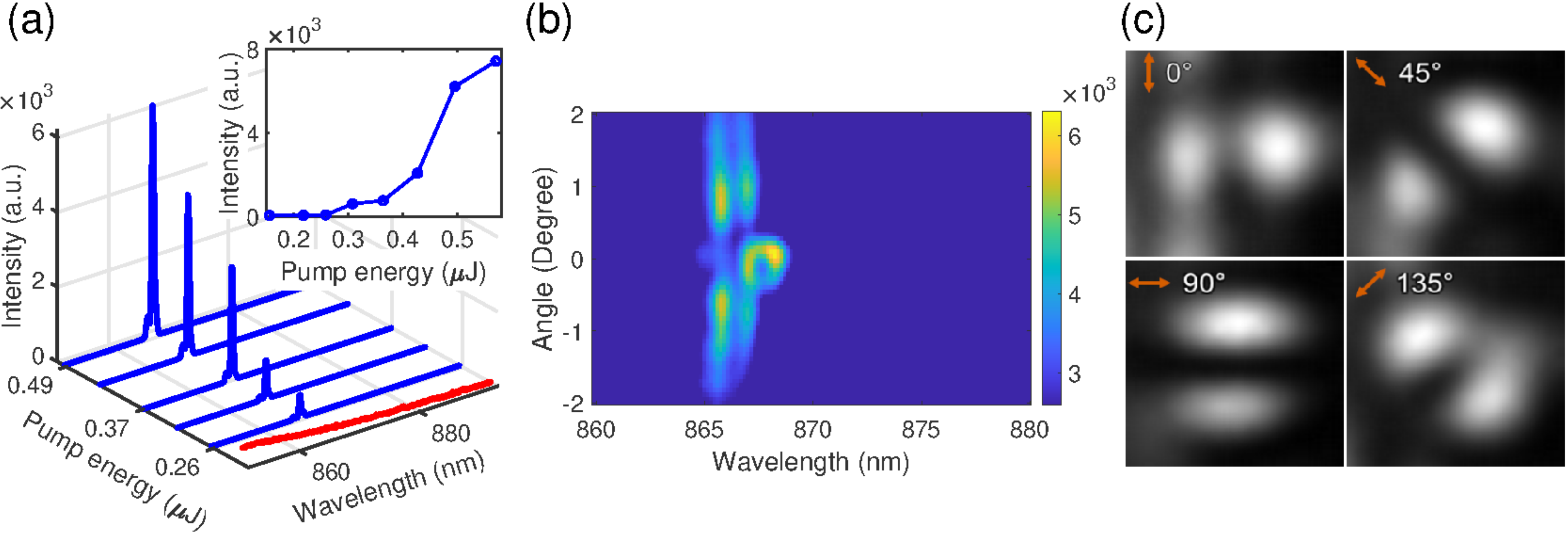}
  \caption{Emission characteristics for p = 550 nm, d = 330 nm  sample. (a) Emission spectra as a function of pump energy. Inset: The emission intensity as a function of pump energy. (b) Angle and wavelength resolved emission at pump energy 1.2 times above the lasing threshold. 
  (c) Polarization resolved direct k-space images.}
  \label{vectorTM1-edited}
  \end{figure*}

The emission properties were studied by overlaying PhC slab with optically pumped liquid gain medium containing organic fluorescent  IR-792 molecules in (EG:BA 1:1) solution. 
In Fig. 4 we present the intensity, wavelength, angle (in $\Gamma-X$ direction) and polarization resolved emission data for  p = 550 nm, d = 330 nm  sample. 
In Fig. 4 (a), the emission spectra as a function of pump fluence are presented. The spontaneous emission of IR-792 dominates the spectral features at low pump fluences (red curve), but an abrupt, narrow of emission peak with a 0.3 nm linewidth emerges at higher pump  fluences. The $\Gamma$-point emission intensity (inset) exhibits a highly nonlinear increase with increasing pump fluence. This, together with narrow emission linewidth suggest stimulated emission/lasing of this mode. 

Next, the spectrometer input slit was opened wide enough to collect all the essential features from the back-focal plane of the objective. In particular we aimed for collecting the expected donut-shaped emission beam of BICs completely. The wider entrance slit results in reduced wavelength accuracy, as the final image  at spectrometer CCD becomes a convolution between the intensity distribution and the different wavelengths at the input slit. For a narrow band light source, however, the far field intensity distribution is directly imaged to the CCD. This allows us to simultaneously identify the central wavelength and the angular distribution of different modes. 
Similar technique has been applied to characterizing the contributions from magnetic and electric dipole emission rates from their emission patterns \cite{Dodson2014-sx}. The resulting angle resolved emission spectrum in the lasing regime is presented in Fig. 4 (b). We observe 3 features: 1) two parabolic shapes at around 865 nm, and, a donut shape at 0 degree,  at a slightly longer wavelength. We associate the two parabolic features to the TE mode (see also the $\Gamma-X$ direction dispersion in Fig. 3 a).


The third feature is donut shaped and slightly redshifted from the parabolic shapes, suggesting it could be due to the TE+1 BIC. Due to diverging radiative Q factor at $\Gamma$-point, BIC emission is expected to have negligible intensity at normal incidence to the sample, in agreement with our observation. 

The simulations suggest this BIC has azimuthal polarization distribution around the $\Gamma$-point, and thus can be identified as TE+1, see also Fig. 1 (b). We employ polarization resolved direct k-space (Fourier plane) imaging to experimentally analyze the polarization properties of the emission. At 0 degree, we observe 2 bright horizontally aligned lobes, indicative of TE-type BIC, see Fig. 4 (c). A consecutive rotation of the polarizer counterclockwise in 45 degree steps rotates the bright lobes to counterclockwise by the same amount. Thus, the measurement confirms the  azimuthal polarization pattern with a winding number of 2$\pi$, suggesting a topological charge of +1. This is in full agreement with the predictions from our simulations as shown in Fig. 1 (b-c). 

The resonance frequencies of BICs can be tuned by the periodicity and the resonance of the bright mode can be changed from low to high wavelength side of the diffracted order by tuning the diameter of the holes (compare Figs. 3 (a) and (b)). By utilizing these degrees of freedom, it is possible to 1) bring the other TE-1 BIC to the wavelength range, where sufficient gain is available for stimulated emission processes and 2) remove the spectral overlap with the radiatively damped bright mode. Next it will be shown that with this approach lasing action for each of the 3 remaining BICs can be realized.

\begin{figure*}[ht]
\centering
\includegraphics[width=\textwidth]{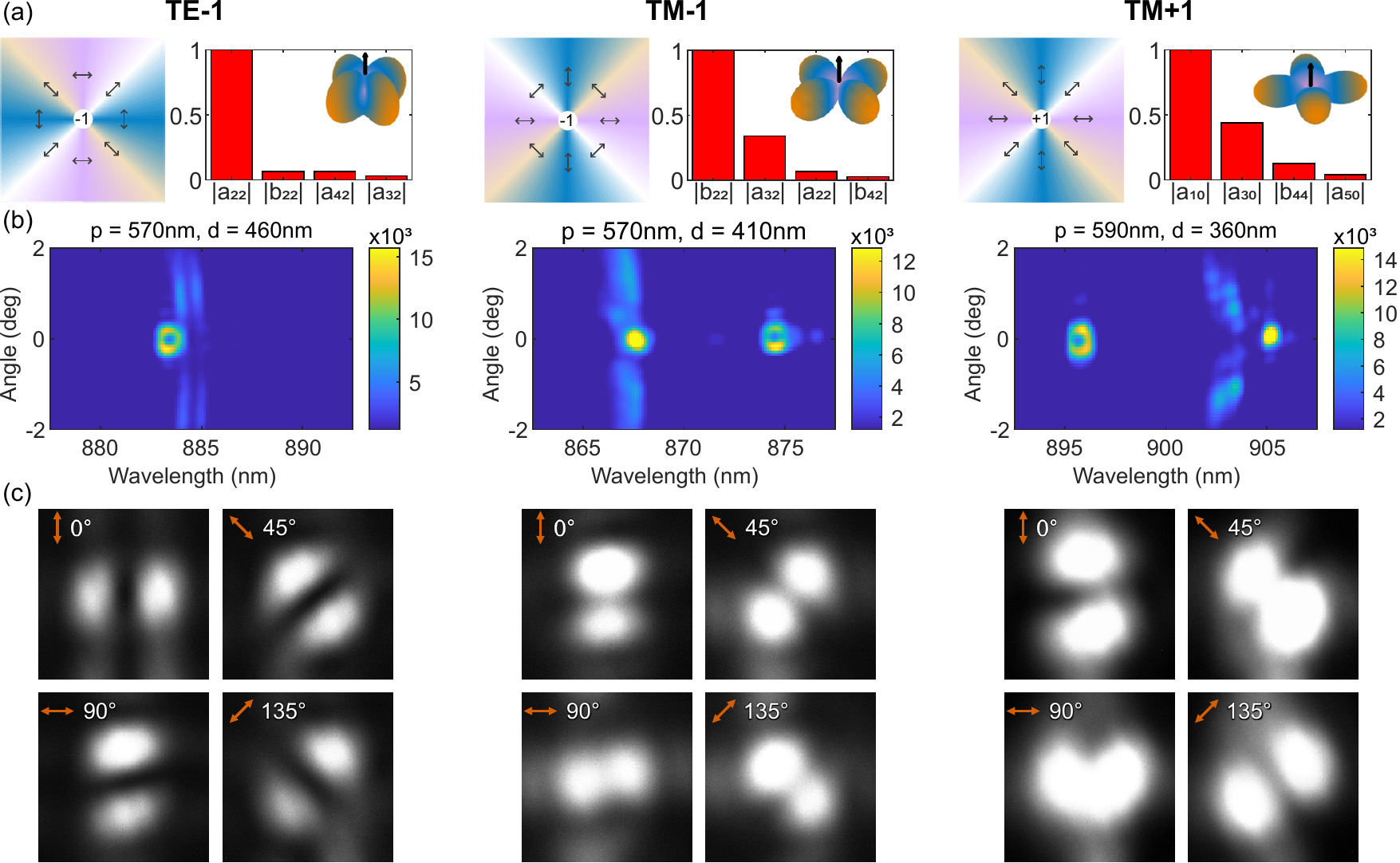}
\caption{Emission characteristics for the TE-1, TM-1 and TM+1 BICs (left, middle and right column, respectively). (a) Multipolar decomposition expansion coefficients(right) and the associated radiation pattern(left) both suggesting the expected topological charge obtained. (b) angle resolved spectra (c) polarization resolved direct k-space imaging for the BICs.}
\label{TM2-edited}
\end{figure*}
  
In Fig. 5 the results for three different samples are presented. First row shows the simulation results, namely the polarization maps and the weights for each BIC obtained from multipolar decomposition theory. Second row shows the angle and wavelength resolved measurement of the lasing signal and the third row represents the
 polarization resolved direct k-space images.
    The left column of Fig. 5 summarizes the emission properties for the sample with p = 570 nm, d = 460 nm. The simulations shown in Fig. 5 (a) indicate a polarization vortex with 2$\pi$ clockwise winding around the $\Gamma$-point, corresponding to topological charge of -1. The near field spherical wave decomposition  recovers the same charge from the index rule \cite{Sadrieva2019}. The experimental angle and wavelength resolved emission distribution shows again 2 parabolic shapes, but with the donut shape intensity distribution now residing on the \textit{short} wavelength side of the parabolic shapes which is expected for the TE-1 BIC from the transmission simulations as well as from the measurements. The direct k-space images in Fig. 5 (c) confirm the clockwise winding of the polarization and the topological charge of -1.

In the middle column of Fig. 5, the emission data for p = 570 nm, d = 410 nm sample is shown. FEM simulation in (a) reveal a clockwise winding of the polarization, suggesting a topological charge of -1, in agreement with the spherical decomposition method. The parabolic shape and the bright mode are  observed on the short wavelength side of the BIC, in agreement with the transmission and simulation data in Fig. 3 (c) for TM-1. Also the direct k-space images shown in the left column of Fig. 5 (c)  confirm the clockwise rotation of the polarization around the $\Gamma$-point.
 
Finally, the TM+1 BIC emission properties are studied with a sample having p = 590 nm d = 360 nm and are summarized in the right column of Fig. 5. The simulations show the expected radial polarization pattern and the spherical wave decomposition coefficients. Both methods suggest a topological charge of +1. The wavelength and angle resolved spectra show a faint parabolic shape at around 902 nm which is associated with the DO, see also Fig. 3. Additionally, a single spot on the high wavelength side of the diffracted order (DO) can be observed (905 nm), which we associate to the bright TM mode  of the PhC slab. Finally, a donut shape pattern at the low wavelength side of the DO (895 nm) can be identified. In the right column of Fig. 5 (c) the polarization resolved k-space image confirms the radial polarization pattern, counter-clockwise winding and therefore the topological charge of +1.

 \begin{figure*}
\centering
 \includegraphics[width=0.8\textwidth]{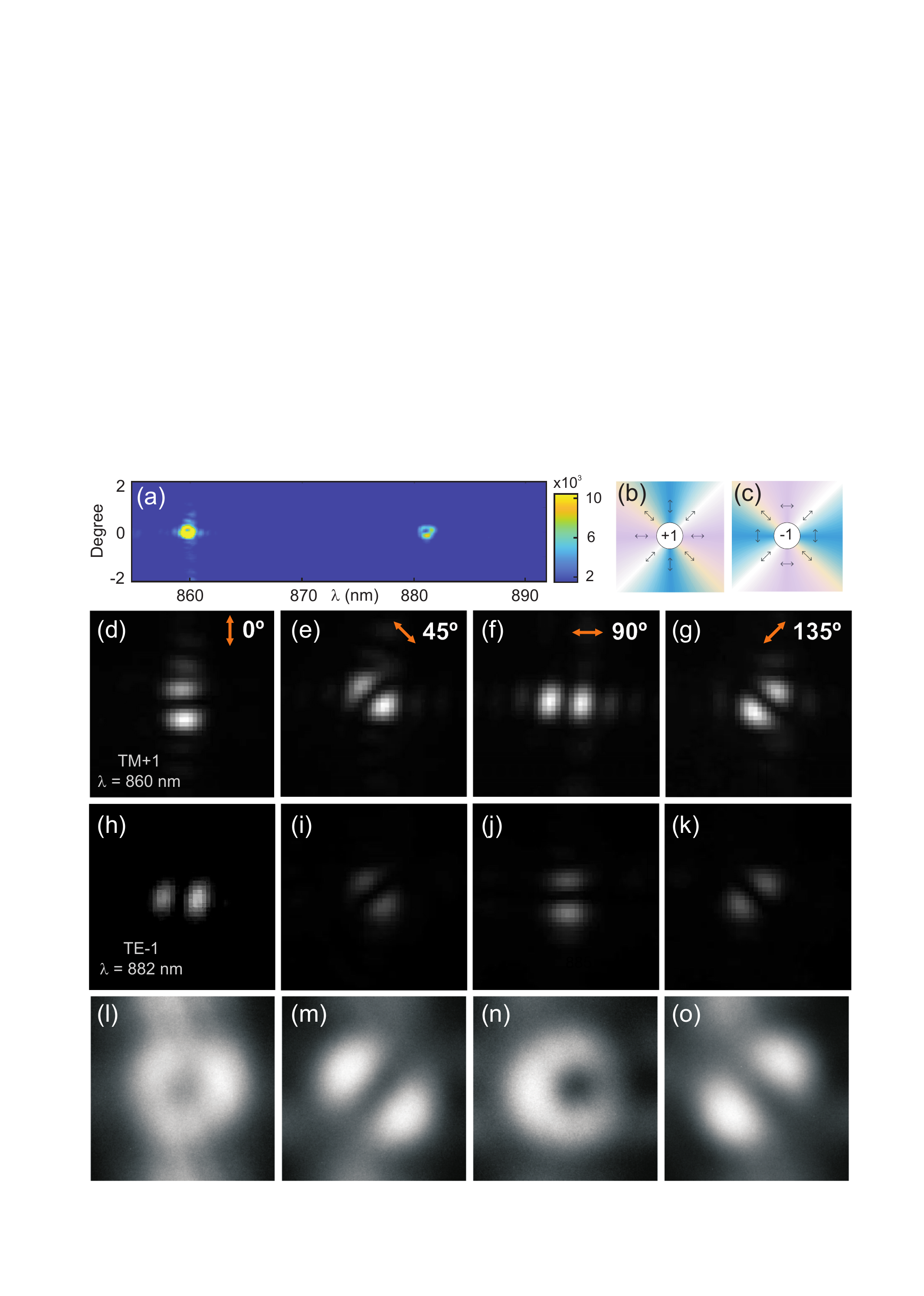}
  \caption{(a) Angle and wavelength resolved emission of p = 560 nm, d = 250 nm sample. (b,c) Polarization maps for the TM+1 and TE-1 BICs obtained from simulations. (d-k) Polarization and angle resolved emission for the TM+1 BIC at 860 nm (d-g), and for TE-1 BIC at 882 nm (h-k). (l-o) Polarization resolved direct k-space images of the beam produced by the 2 coexisting BICs.}
  \label{twodonuts1}
  \end{figure*}
  

Finally, in  Fig. \ref{twodonuts1}, we present an interesting case for a sample p = 560 nm and d = 250 nm, where 2 non-degenerate BICs exhibit lasing. Angle resolved spectra in Fig. 6 (a) reveal BICs at 860 and 882 nm, whose polarization properties are shown in Figs. 6 (b) and (c), respectively. As both BICs lase at the $\Gamma$-point of the lattice, their emissions overlap in the far field. In order to recover the polarization of the BICs, we employed polarization, angle and wavelength resolved measurement scheme by inserting a rotatable linear polarizer before directing the beam to the spectrometer. In Figs. 6 (d-g) are  shown the emission patterns of the BIC at 860 nm with different polarizer angles (0, 45, 90, 135 degrees). Notably, the bright lobes are aligned vertically for the 0 degree case (d), indicative of a TM related BIC. As the polarizer is rotated counterclockwise, the lobes rotate counterclockwise as well, suggesting a topological charge of +1 (h-k). The BIC at 882 nm, on the other hand, exhibits 2 horizontally located lobes at 0 degree polarizer angle (h), which rotate clockwise indicating TE related BIC with a topological charge of -1. A typical far field intensity pattern for the coexisting TM+1 and TE-1 BICs are shown in Fig. 6 (l-o). Intriguingly, at 0 degree, a somewhat even intensity pattern is observed around the perimeter of the donut shape beam. At 45 and 135 degrees  of polarizer (m and o), the emission pattern displays 2 bright lobes at 45 and 135 degree angles, respectively. However, 90 degree angle produces an even intensity distribution similar to (l). These observations can be rationalized by detailed inspection of the polarization properties of the 2 BICs, see Fig. 6 (b) and (c). At 0 and 90 degree angles, one BIC produces two horizontal bright lobes while the other one produces two vertical lobes. This results to  almost even intensity distribution around the perimeter of the donut. However, with 45 and 135 degrees, both BICs produce bright lobes at 45 and 135 degrees, respectively. 


\section{Conclusions}

By tuning the lattice parameters, namely diameter and periodicity of PhC slab, we have demonstrated lasing action from each of the four BICs enabled by the C$_{4v}$ symmetry of the square lattice.                                
The topological charge and type of each BIC were experimentally determined with polarization resolved k-space imaging. The polarization states as well as topological charges of each BIC were obtained from FEM simulations with two complementary methods, the first one employing k-space polarization analysis in the infinite lattice, and the second one based on multipolar decomposition analysis of the microscopic polarization currents in the unit cell of the PhC. The predictions obtained from FEM simulations fully agree with the observations in each case. For the sample exhibiting two non degenerate BICs, a polarization, angle and wavelength resolved hybrid method was utilized for the analysis.

We believe our results enable a priori design of topological lasers. Interesting future  prospects include realizing  topological charges of for instance $q = \pm (2,3,4...)$ by similar approach. This should be feasible by utilizing higher frequency (2nd, 3rd, 4th) diffracted orders of the lattice. By modification of the band structure, lasing with fractional topological charges may be possible \cite{PhysRevB.99.180101}.
Obtaining complete control of the topological charge and the  polarization state of the associated laser light could be appealing in various contexts, such as nanoscale particle trapping and manipulation, all optical information processing as well as as well as gain assisted metamaterials in general. 

Yet another interesting prospect is related to the coexistence of two lasing BICs with opposite topological charges. Superpositions of vector beams have been studied in various contexts, in particlular related to optical trapping. Typically, vortex beams are created by external components outside the laser cavity, and therefore the beams have the same frequency. Here, the feedback in the PhC slab creates two non degenerate BICs, whose temporal coexistence would imply complex evolution of polarization states, and intensity variation at the beat note frequency.

Utilization of the appealing features of BICs in several contexts, including the control of polarization and topological charge of lasing action, nanoscale particle trapping and manipulation, as well as as well as gain assisted metamaterials could benefit from the present study.

\section*{Methods}

\textbf{Sample fabrication.} A 170-nm-thick Si$_3$N$_4$ layer was grown on a SiO$_2$ substrate by plasma-enhanced chemical vapour deposition (PECVD), and then spin-coated with a layer of positive electron-beam resist (PMMA950K, A4) and an additional layer of conductive polymer (AR-PC 5090.02). The periodic photonic crystal (PhC) pattern was fabricated onto the PMMA layer using electron-beam lithography (ZEISS sigma 300) and subsequently transferred to Si3N4 layer by reactive-ion etching (RIE). SF$_6$/O$_2$ gas was used to etch Si3N4 and the etched depth could be controlled by etching time. The etched depth of Si$_3$N$_4$ layer was about 120~nm.

\textbf{Simulation.} The band structure of PhC slab is simulated using a finite-element method (COMSOL MULTIPHYSICS). The meshing resolution is set to ensure the accuracy of electric and magnetic field calculations. The eigenvalue solver is used to get the information of polarization states and $Q$-factors. We can extract the information of the polarization states denoted as the major axis of polarization states~\cite{yariv2006photonics,mcmaster1954polarization}, which can be calculated through the Stokes parameters: 
\begin{align*}
  S_0  
  &= |E_x|^2 + |E_y|^2,\\
  S_1  
  &= |E_x|^2 - |E_y|^2,\\
  S_2  
  &= 2 \rm{Re}(E_x E_y^*),\\
  S_3  
  &= -2 \rm{Im}(E_x E_y^*).
\end{align*}
Then the polarization azimuthal angle expressed as $Arg(S_1+i S_2)/2$ can be easily obtained. The eigenvalue solver returns complex eigenfrequencies $f$, and the corresponding radiative $Q$-factor can be calculated as $Q=\rm {Re}(f)/ \rm{Im}(f)$. 

The transmission spectra are calculated by using rigorous coupled-wave analysis (diffract mode, RSOFT), a commercially available software~\cite{moharam1981rigorous,moharam1995formulation}. The wavelengths of incident light range from 830~nm to 970~nm in steps of 0.1~nm and the incident angle varies from $-4^\circ$ to $4^\circ$ in steps of $0.1^\circ$. Two harmonics are tested for the convergence of simulation results. In all simulations, the refractive indices of  Si$_3$N$_4$ and SiO$_2$ are set as 1.975 and 1.45, respectively.

\bibliography{main}

\noindent\textbf{Acknowledgements}
This work is part of the Academy of Finland Flagship Programme, Photonics Research and Innovation (\textrm{PREIN}), decision 320166.

\noindent\textbf{Funding}
Academy of Finland Flagship Programme, Photonics Research and Innovation (\textrm{PREIN}) 320166; Academy of Finland (322002).


\end{document}